# Chiral Topological Semimetal with Multifold Band Crossings and Long Fermi arcs

Niels B. M. Schröter[1*], Ding Pei[2], Maia G. Vergniory[3,4], Yan Sun[5], Kaustuv Manna[5], Fernando de Juan[3,4,6], Jonas. A. Krieger[1,7,8], Vicky Süss[5], Marcus Schmidt[5], Pavel Dudin[9], Barry Bradlyn[10], Timur K. Kim[9], Thorsten Schmitt[1], Cephise Cacho[9], Claudia Felser[5], Vladimir N. Strocov[1], and Yulin Chen[2*]

[1]*Swiss Light Source, Paul Scherrer Institute, CH-5232 Villigen PSI, Switzerland*
[2]*Clarendon Laboratory, Department of Physics, University of Oxford, Oxford OX1 3PU, United Kingdom*
[3]*Donostia International Physics Center, 20018 Donostia-San Sebastian, Spain*
[4]*IKERBASQUE, Basque Foundation for Science, Maria Diaz de Haro 3, 48013 Bilbao, Spain*
[5]*Max Planck Institute for Chemical Physics of Solids, Dresden, D-01187, Germany*
[6]*Rudolph Peierls Centre for Theoretical Physics, University of Oxford, Department of Physics, Clarendon Laboratory, Parks Road, Oxford, OX1 3PU, United Kingdom*
[7]*Laboratory for Muon Spin Spectroscopy, Paul Scherrer Institute, CH-5232 Villigen PSI, Switzerland*
[8]*Laboratorium für Festkörperphysik, ETH Zurich, CH-8093 Zurich, Switzerland*
[9]*Diamond Light Source, Didcot, OX110DE, United Kingdom*
[10]*Department of Physics and Institute for Condensed Matter Theory, University of Illinois at Urbana-Champaign, Urbana, IL, 61801-3080, USA*
*Correspondence to: niels.schroeter@psi.ch , yulin.chen@physics.ox.ac.uk

**Topological semimetals in crystals with a chiral structure (which possess a handedness due to a lack of mirror and inversion symmetries) are expected to display numerous exotic physical phenomena, including fermionic excitations with large topological charge[1], long Fermi arc surface states[2,3], unusual magnetotransport[4] and lattice dynamics[5], as well as a quantized response to circularly polarized light[6]. To date, all experimentally confirmed topological semimetals exist in crystals that contain mirror operations, meaning that these properties do not appear. Here, we show that AlPt is a structurally chiral topological semimetal that hosts new fourfold and sixfold fermions, which can be viewed as a higher spin generalization of Weyl fermions without equivalence in elementary particle physics. These multifold fermions are located at high symmetry points and have Chern numbers larger than those in Weyl semimetals, thus resulting in multiple Fermi arcs that span the full diagonal of the surface Brillouin zone. By imaging these long Fermi arcs, we experimentally determine the magnitude and sign of their Chern number, allowing us to relate their dispersion to the handedness of their host crystal.**

An object that cannot be superimposed with its mirror image is said to be chiral, a concept first proposed by Lord Kelvin that has found widespread applications across the modern sciences, from high energy physics to biology. In condensed matter physics (where the properties of crystalline materials are tightly constrained by spatial lattice symmetries), chiral crystals can only be found in the 65 Sohncke space groups, which contain only orientation-preserving operations, and can therefore be assigned a handedness. The structural chirality of these systems can endow them with fascinating properties, such as natural optical activity, negative refraction in metamaterials[7], non-reciprocal effects such as magnetochiral birefringence of light or electronic magnetochiral anisotropy[8,9], chiral magnetic textures such as helices and skyrmions[10], or unusual superconductivity[11]. In the recently discovered Weyl-semimetals, not just the crystal structure itself, but also the electronic wavefunctions can exhibit a chirality at point-like two-band crossings of the quasiparticle dispersion, which are known as Weyl fermions[12–15]. These crossings are topologically protected because they carry a topological charge, a quantized Berry flux through any surface enclosing them in momentum space. This charge has integer magnitude C= ±1, which is known as the Chern number and gives the electronic wavefunction a handedness. This topological property results in a plethora of exotic phenomena, such as Fermi-arc surface states[16], unconventional magnetoresistance[12,17,18], nonlocal transport[19], and many other nonlinear and anomalous responses[20–25]. While relativistic Weyl fermions must obey Lorentz symmetry, condensed matter systems are free from this constraint and instead need only be invariant under the space group of the crystal. This freedom gives rise to a much richer variety of topological semimetals, such as tilted Weyl semimetals[26], and topological semimetals involving three, four, or six band crossings protected by symmetry[1,27,28]. When such multifold crossings are realized in chiral crystal structures, the effective Hamiltonian near the crossings is a higher spin generalization of the Weyl Hamiltonian H=**k·S.** Here, **k** is the crystal momentum and the matrices **S** are adiabatically connected to spin matrices for spin-1, spin-3/2 or the direct sum of two spin-1 copies for the threefold, fourfold and sixfold crossings, respectively[1]. The bands of these multifold fermions feature higher Chern numbers (which also bestows their wavefunction with a handedness), and therefore host more Fermi arcs and chiral Landau levels (which are also qualitatively different[1]) compared to Weyl semimetals, which enhances many of the phenomena associated with Weyl physics. Spin 1-fermions, in particular, are markedly different from the triple band crossings recently discovered in WC[29] and MoP[30], which are not structurally chiral and feature no well-defined Chern numbers.

A nonzero Chern number does not require a chiral crystal structure, but topological semimetals that also possess structural chirality have many unique properties[31] (see Fig. 1A for an overview), such as the gyrotropic magnetic effect[4], a finite frequency version of the chiral magnetic effect first predicted for the quark-gluon plasma[32]; the signatures of the chiral anomaly that arise in phonon dynamics[5] or the magnetochiral anisotropy[9]; and most notably the quantized circular photogalvanic effect[6,33], the only experimental probe that can directly measure the Chern numbers of a topological semimetal. However, all experimentally confirmed topological semimetals – of the Weyl type or any generalization – have been synthesized so far only in non-chiral crystal structures. Orientation reversing operations such as mirror symmetries in these materials require the existence of Weyl nodes with opposite Chern numbers at the same energy (see Fig. 1B), which force the aforementioned properties to vanish. In a chiral topological semimetal (i.e. a semimetal with nonzero Chern number and broken mirror and inversion symmetry), nodes of opposite Chern number are generically separated in energy. The absence of a viable material to observe these properties has severely limited progress in the field up to now, despite an extensive theoretical search for them[31]. Promising material candidates that could display these phenomena were recently predicted in a family of transition-metal silicides and platinum and palladium alluminides[1–3], which crystallize in the chiral space group 198. In this family of materials, a spin-3/2 fermion is realized near the Fermi level at the $\Gamma$ point (which corresponds to the elusive Rarita-Schwinger fermion predicted in particle physics), while a double spin-1 fermion is realized at the R point. The Chern number count predicts at least four chiral surface Fermi arcs that traverse the diagonal of the surface Brillouin zone, thus connecting the projections of the $\Gamma$ and R points, which would span the largest portion of the Brillouin zone of any topological material. In addition, the nodes at $\Gamma$ and R have a significant energy separation, which is a practical advantage to realize the physical phenomena discussed above. Since sample surfaces of the silicides have been proven difficult to prepare by in-situ crystal cleaving, we chose to study the material AlPt.

In this work, we conducted angle-resolved photoelectron spectroscopy (ARPES) measurements of AlPt crystals at low photon energies (VUV-ARPES, hv < 150 eV) to investigate their surface electronic structures, as well as soft X-ray energies (SX-ARPES, hv > 300 eV) to study their bulk band structures. To understand the experimental results, we also performed *ab-initio* band structure calculations (see the methods section for more details).

AlPt crystallizes in the cubic chiral B20 structure of FeSi that can be considered as a distorted rocksalt structure where the atomic positions are displaced along the (111) direction with a lattice constant of a=4.863. Since the material is not layered, it can be cleaved along different Miller planes, such as (001), (110) and (111), as is illustrated in Fig. 1C. Its cubic Brillouin zone is shown in Fig. 1D, which also illustrates the fourfold and sixfold crossings at the Γ and R points that were predicted in Refs. 1,2, as well as the Brillouin zone boundary for planes parallel to the (111) and (110) Miller planes. We confirmed its crystal structure and elemental composition via X-ray powder diffraction and core level photoemission spectroscopy, respectively; the results are shown in Fig. 1E (more details about the samples, including information about the presence of a small impurity phase, can be found in the supplementary materials). The experimental Fermi surfaces measured on the (110) and (111) surface are displayed in Fig. 1F and 1G, respectively, showing agreement with the expected rectangular and trigonal Brillouin zone boundary. Fig. 1G also displays the experimental band dispersion near the R point, clearly showing the expected band crossing at R - the first indication for the existence of sixfold fermions in AlPt.

To further investigate these exotic new fermions, we performed a detailed comparison between the experimental bulk band structure obtained with SX-ARPES and our *ab-initio* calculations, which is shown in Fig. 2. We focus here primarily on data obtained from the (111) surface, while further results on other surfaces can be found in the supplementary materials. Fig. 2A-D display data obtained with a photon energy of hv=554 eV, which corresponds to a measurement plane in the Brillouin zone that contains the R and X points, as shown by the orange dashed line in Fig. 1D and 2C(ii), while Fig. 2E was obtained with hv=680 eV to access Γ and M points (details about the determination of the measurement planes along $k_z$ is given in the supplementary materials). The experimental dispersion along X-R-X shown in Fig. 2A(i) is in qualitative agreement with our *ab-initio* calculations shown in Fig. 2A(ii), which highlights that the experimentally observed nodal point consists of a sixfold crossing of three doubly degenerate bands (the degeneracy is lifted along other momentum directions). The observed position of the nodal point at binding energy $E_b \approx 0.77$ eV is close to the calculated value at binding energy $E_c \approx 0.69$ eV. The experimental constant-energy surfaces displayed in Fig. 2B show that this node exists as a band crossing along all in-plane momentum directions and that the bands involved in the sixfold crossing disperse almost linearly towards the Fermi level, leading to an electron pocket at R. The experimental Fermi surface in Fig. 2C(i) and constant-energy surface in Fig.

2D(i) are also in qualitative agreement with the calculations shown in Fig. 2C(ii) and 2D(ii), respectively, which confirms the position of the sixfold fermion at R and reflects the symmetry expected from this measurement plane. Small quantitative deviations (e.g. the apparent absence of the predicted small hole pockets in the experimental data) are due to matrix element effects in the photoemission process not included in the calculations, which do not influence our conclusion about the existence of the sixfold Fermion at R. Fig. 2E(i) shows the dispersion along the M-Γ-M plane, which is also in excellent agreement with the calculation shown in Fig. 2E(ii), and indicates the expected fourfold fermion at Γ. Overall, the agreement between our SX-ARPES data with our *ab-initio* calculations clearly establishes the existence of fourfold and sixfold fermions in AlPt.

To investigate the topological nature of these fermions, we compare in Fig. 3 the experimental surface electronic structure of the (001) cleavage plane, which is expected to host topological surface Fermi arcs, with the respective experimental bulk electronic structure and ab-initio calculation. Fig. 3A shows the position of the bulk Fermi surface pockets in the $k_z=\pi/a$ plane measured with SX-ARPES, which suppresses contributions from surface states due to the large probing depth (also see Fig. S2 B for the bulk FS for $k_z=0$). These bulk pockets are well described by the bulk DFT calculations shown in Fig. 3B, where the measurement plane of Fig. 3A is indicated by the red surface (the experimental data also includes contributions from neighboring planes due to $k_z$ broadening). We can see from the projections of the bulk calculations to the (001) surface (shown as blue contours on top of Fig. 3B) that the gaps between the bulk pockets shown in Fig 3A remain gapped for all $k_z$ values (i.e. they are projected bulk gaps), meaning that no bulk bands are expected to appear in those gaps at any photon energy (since changing the photon energy would correspond to probing bands at different $k_z$ momenta). When comparing our surface sensitive VUV-ARPES data shown in Fig. 3C to the bulk data presented in Fig. 3A, we can identify large S-shaped Fermi arcs that are threading through these projected bulk gaps and are connecting the projected $\bar{\Gamma}$ and $\bar{R}$ points (highlighted by red dashed lines as guide to the eye), which are spanning the entire diagonal of the surface Brillouin zone and must originate from the surface. This conclusion is further corroborated our ab-initio slab calculations that are also shown in Fig. 3C (showing surface spin-up and spin-down bands as red and blue lines, respectively, whilst bulk bands are shown in white), and projected bulk band contours (solid blue lines), which match our experimental data.

To further enhance the contrast of the S-shaped Fermi arcs, we show a curvature plot[34] of the experimental Fermi surface in Fig. 3E, which also clearly indicates the S-shaped Fermi arcs. Additionally, we also display the experimental energy vs. momentum dispersion of experimental data along the $\bar{R}$-$\bar{X}_1$ direction in Fig 3D (i), which also clearly shows the Fermi arc bands dispersing in the projected bulk band gap, and which are reproduced by the ab-initio slab calculations in Fig. 3D (ii). Due to spin-orbit coupling the surface Fermi arcs must be spin-split, which is corroborated by the band splitting in our slab calculations. We therefore conclude that that our surface sensitive VUV-ARPES data contrasted with our bulk sensitive SX-ARPES data and ab-initio calculations provides strong evidence for the existence of two surface Fermi arcs that are approaching $\bar{\Gamma}$ and $\bar{R}$, and which must be spin-split due to spin-orbit coupling, which means that the new multifold Fermions R and Γ must have a Chern number with magnitude |C|=4.

A further analysis of the Fermi arc dispersion can not only determine the magnitude of the Chern numbers in AlPt, but also their sign, which is intimately related to the handedness of the crystal structure. As we illustrate in Fig. 4 A-B, the two enantiomers of AlPt can be distinguished by the handedness of the helix that is formed by their Al atoms along the (111) direction (which can be left-handed or right-handed), and by the configuration of the signs of their Chern number (either [Γ+/R-] or [Γ-/R+]). The fact that the Chern number signs are inverted between the two enantiomers can be understood by the fact that the integral of the Berry curvature acquires a minus sign upon a mirror operation since it involves the curl of the Berry connection. Which of these two sign configurations is realized for a particular enantiomer is not fixed by symmetry but can be calculated by ab-initio calculations. Here we find that the left-handed Al-helix corresponds to a positive Chern number at Γ, which means that the Berry curvature flows from Γ to R, which in turn defines the direction along which the Fermi arcs disperse for line cuts between $\bar{\Gamma}$ and $\bar{R}$: They become right-moving for a cut to the left of $\bar{\Gamma}$ and left-moving for a cut to the right of $\bar{\Gamma}$. This directionality becomes reversed in the other enantiomer since the flow of Berry curvature is reversed. By examining our experimental data (shown in Fig. 4 C-E) along the two line cuts to the left and right of $\bar{\Gamma}$ that are passing through the projected bulk band gaps of AlPt (indicated by red arrows), we find that the Chern number sign configuration in the sample measured here was [Γ+/R-], and that the corresponding crystal structure involved a left-handed Al helix. To the best of our knowledge, such a connection between surface state dispersion and crystalline chirality was not made before and can be used in the future to experimentally determine the handedness of a chiral topological semimetal.

# Figures

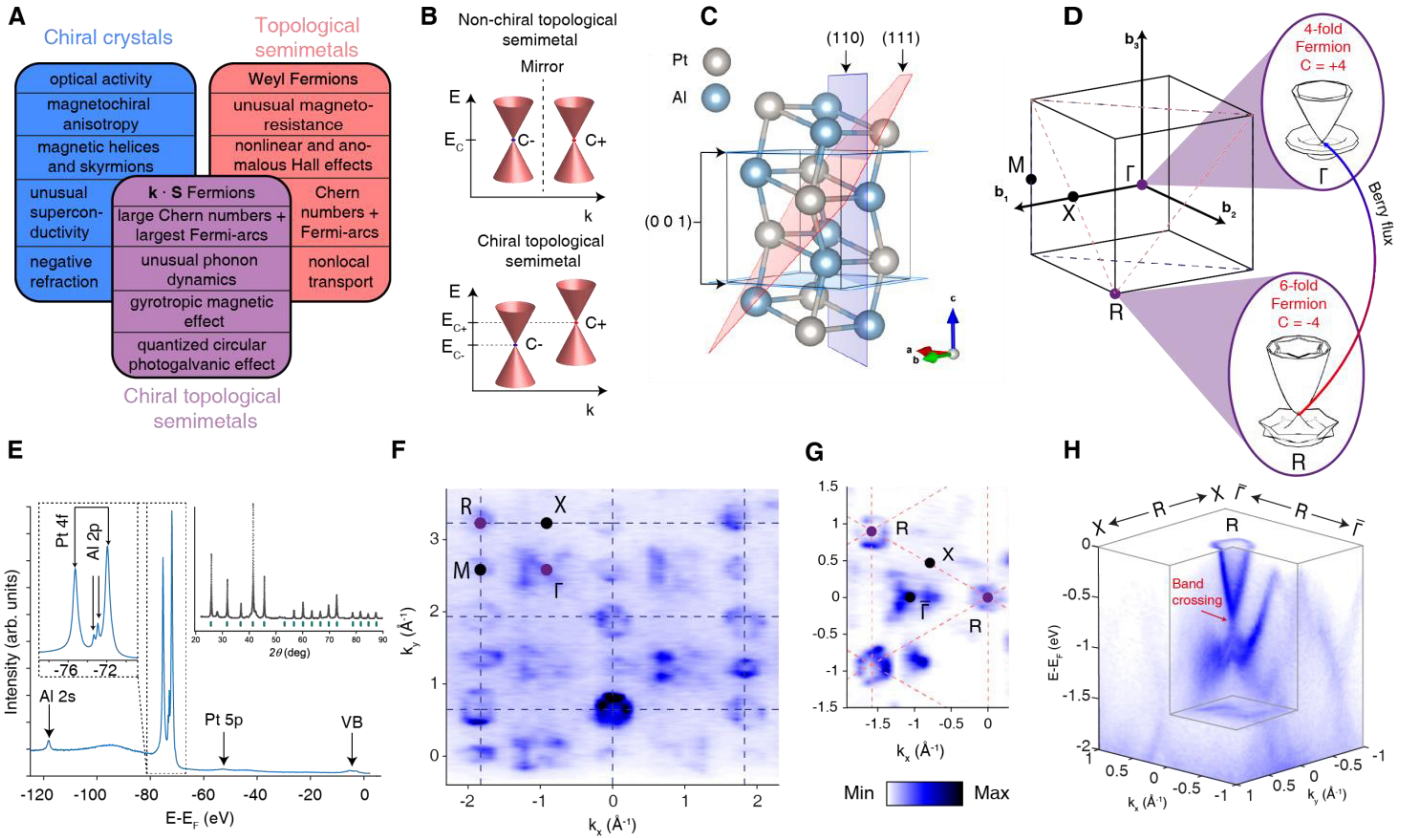

**Fig. 1**. **Basic characteristics of chiral topological semimetals and AlPt.**
(**A**) Overview of various exotic phenomena in chiral crystals, topological semimetals, and chiral topological semimetals. See also Ref. 31. (**B**) Energy positions of band crossings with opposite Chern number ($C_{\pm}$) in non-chiral topological semimetals (top) and chiral topological semimetals (bottom). (**C**) AlPt crystal structure and possible cleavage planes. (**D**) The cubic Brillouin zone of AlPt, featuring an illustration of fourfold and sixfold fermions. (**E**) Experimental core level spectrum and powder diffraction pattern (inset – green marks indicate calculated Bragg peaks for AlPt). (**F**) Experimental Fermi surface of the (110) cleavage surface, including the boundary of the bulk Brillouin zone, taken at hν = 700 eV with right-circular polarization. (**G**) Experimental Fermi surface of the (111) cleavage surface, taken with hν=800 eV with left-circular polarization. (**H**) 3D intensity plot of our ARPES data measured with hν=554 eV and linear-vertical polarization, showing a band crossing at the R point.

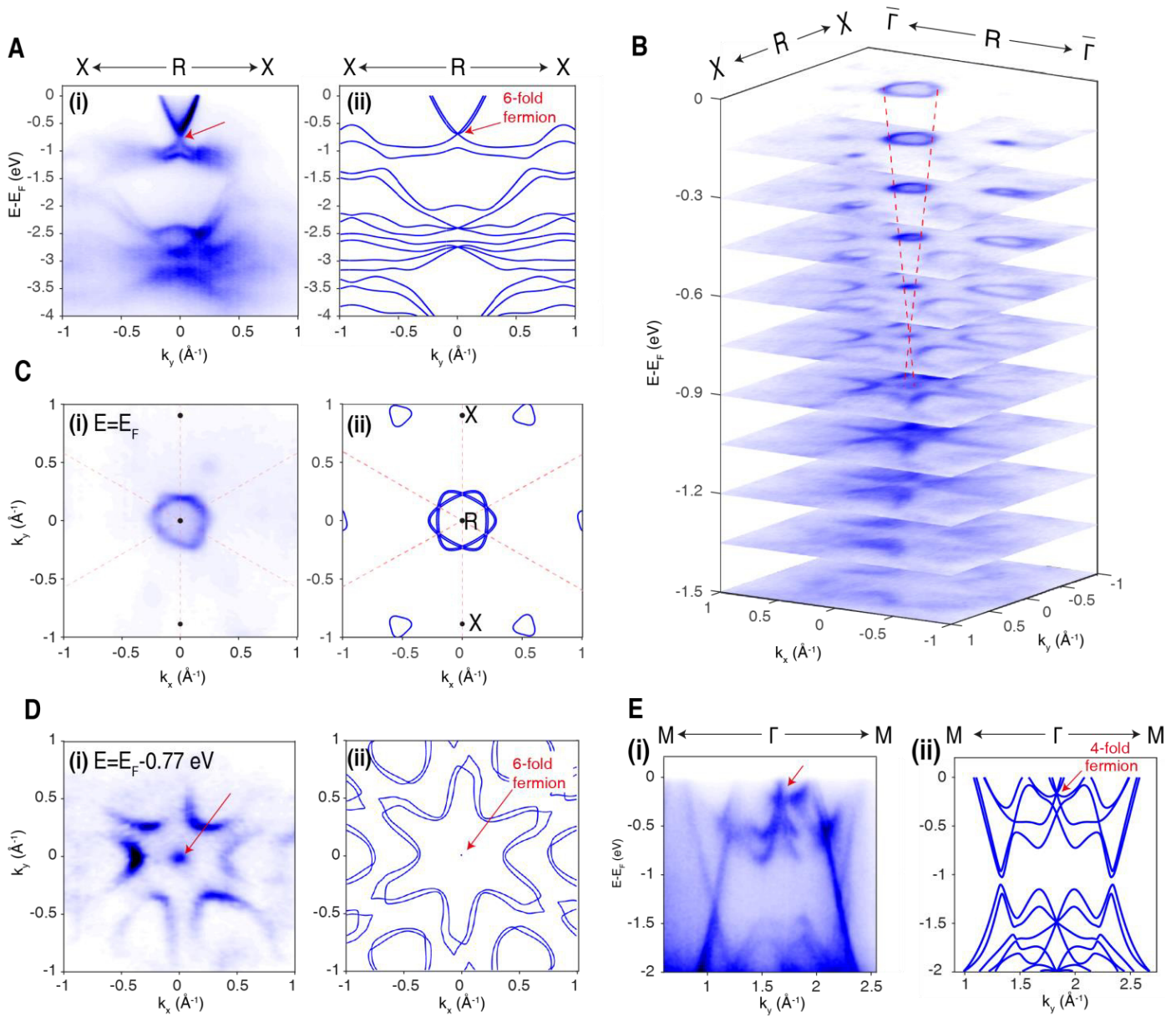

**Fig. 2. Bulk electronic structure of AlPt.**
(**A**) Comparison between experimental and calculated band dispersion along X-R-X, the red arrow indicates the sixfold band crossing, measured with hν=554 eV. Note that all bands along this high symmetry direction are doubly degenerate. (**B**) Energy evolution of the constant-energy surfaces at the R point, taken with hν=554 eV. Red dashed lines are a guide for the eye. (**C-D**) Comparison between experimental and calculated Fermi- and constant-energy surface (which are shown for binding energy $E_b$ = 0.77 eV for the experimental data and $E_c$ = 0.69 eV for the calculated data), measured with hν=554 eV. The calculation shows additional hole pockets at the Fermi level that are suppressed in the experiment due to matrix element effects in the photoemission process. (**E**) Comparison between experimental and calculated band dispersion along M - Γ -M, the red arrow indicates the fourfold band crossing, taken with hν=680 eV. All experimental data was measured with linear-vertical polarization.

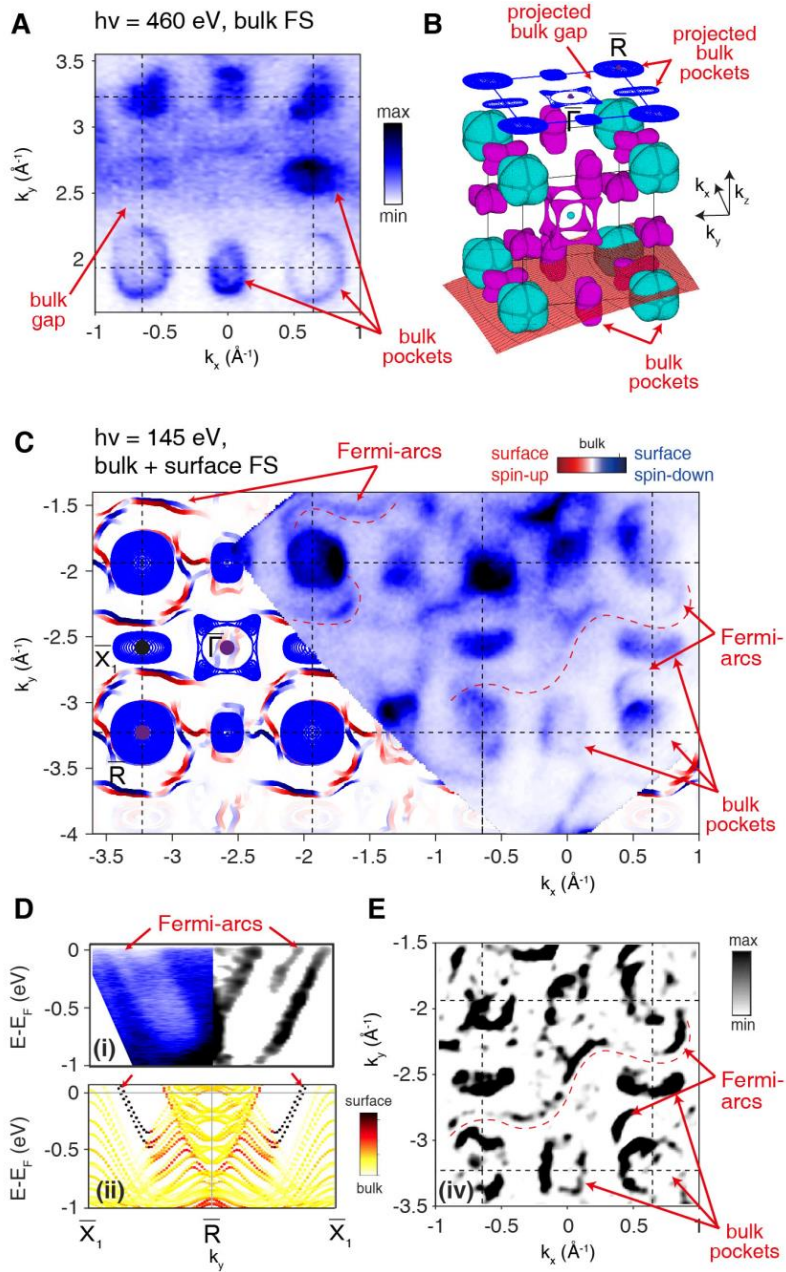

**Fig. 3: Bulk electronic structure and surface Fermi arcs on the (001) cleavage plane.**
(**A**) Bulk Fermi surface probed at hv=460 eV in the $k_z = \pi/a$ plane. (**B**) Calculated 3D bulk Fermi surface of AlPt (hole pockets in magenta and electron pockets in cyan). The blue contours on top show the projection of the bulk pockets to the (001) surface Brillouin zone. The red surface on the bottom approximately corresponds to the measurement plane at hv=460 eV shown in (A). (**C**) Experimental Fermi surface map of measured with hv=145 eV and linear-horizontal polarization, integrated over 100 meV at the Fermi level. Also shown are ab-initio slab calculations (blue and red lines), as well as the projected bulk calculations (blue contours). Both the experimental data and the slab calculation show surface Fermi arcs threading through the projected band gaps and spanning the full diagonal of the surface Brillouin zone (red dashed lines are a guide for the eye). (**D, i**) Experimental band dispersion along the $\bar{X}_1 - \bar{R}_1$ direction, raw data on the left and curvature plot on the right. (**D, ii**) Corresponding ab-initio slab calculation. (**E**) Curvature plot of the experimental data shown in (C), red dashed line is guide to the eye for the Fermi arc.

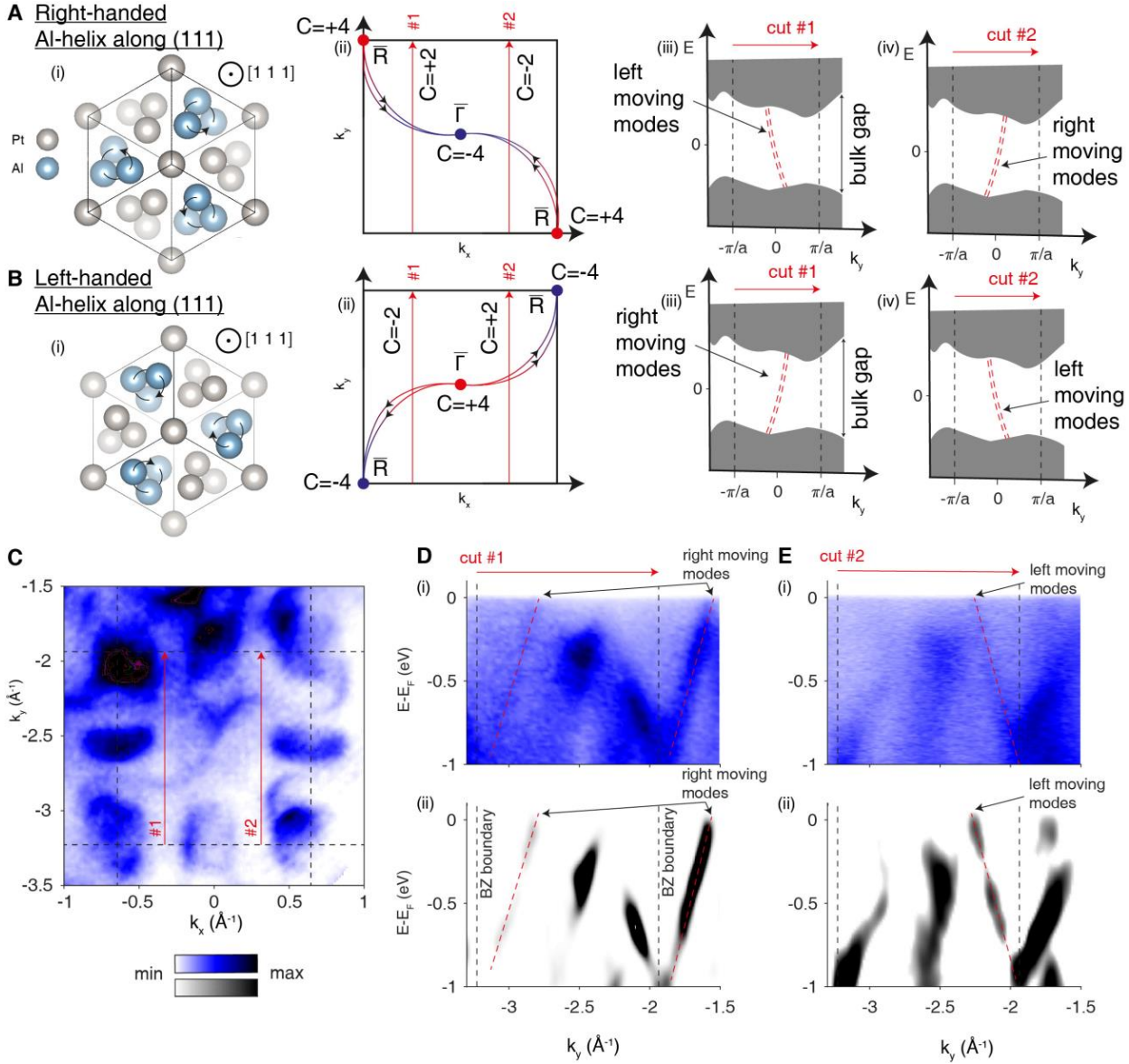

**Fig. 4: Connection between the directionality of the surface Fermi arcs and the handedness of the bulk crystal. (A-B)** (i) Crystal structures of two enantiomers (mirror images) of AlPt, which can be distinguished by the handedness of the helix formed by Al atoms along the (111) direction. (ii) Corresponding configuration of the Chern number signs, Fermi arc dispersion, and the flux of the Berry curvature (indicated by arrows and red/blue color, flowing from positive to negative Chern number). (iii-iv) Dispersion of the corresponding Fermi arcs along the line-cuts shown as red arrows in (ii). **(C)** Zoom-in of the experimental data shown in Fig. R1 C, red arrows show position of the two line cuts shown in subfigures D-E. **(D-E)** Two line cuts to the left and to the right of the $\bar{\Gamma}$, point, respectively. (i) shows the raw data and (ii) shows the curvature plot. Red dashed lines are guide to the eye of the dispersion of the surface Fermi arcs. Black dashed lines indicate the boundaries of the (001) surface Brillouin zone.

**Author contributions:**

N.B.M.S. conducted the SX-ARPES experiments with the support of J.A.K. and V.N.S. and the VUV-ARPES experiments with the support of D.P. and P.D. The experimental data was analyzed by N.B.M.S. and D.P. M.G.V. and Y.S. performed the VASP slab and bulk ab-initio calculations, and N.B.M.S. performed the Wien2k bulk calculations with the support of D.P. F.J. and B.B. provided further theoretical support. K.M., V.S., and M.S. grew the samples and K.M. performed the powder XRD refinement. P.D., T.K.K., T.S., C.C., and V.N.S. maintained the ARPES end stations. N.B.M.S. and F.J. wrote the manuscript with input and discussion from co-authors. V.N.S., C.F., and Y.C. supervised the research.

**Acknowledgements:**

We are thankful for excellent technical support from L. Nue and A. Pfister. We acknowledge Diamond Light Source for access to beamline I05 (proposals no. SI19883 and SI21400). We acknowledge the Paul Scherrer Institut, Villigen, Switzerland for provision of synchrotron radiation beamtime at beamline ADRESS of the SLS. N.B.M.S. acknowledges partial financial support from Microsoft. Y.L.C. acknowledges support from the Engineering and Physical Sciences Research Council Platform Grant (Grant No. EP/M020517/1). D.P. acknowledge the support from China Scholarship Council. F.J. acknowledges funding from the European Union's Horizon 2020 research and innovation programme under the Marie-Sklodowska Curie grant agreement No. 705968. J.A.K. acknowledges the Swiss National Science Foundation (SNF-Grant No. 200021_165910). Part of the work of B.B. and M.G.V. was carried out at the Aspen Center for Physics, which is supported by National Science Foundation grant PHY-1607611. M.G.V. was supported by IS2016- 75862-P national project of the Spanish MINECO. K. M. and C. F. acknowledge financial support from ERC through Grant No. 742068 "TOP-MAT".

**Methods:**

Sample growth

The polycrystalline sample of AlPt was first prepared by arc melting the stoichiometric amount of the high pure Al and Pt metals. Then the ingot was crushed into fine powder and the final sample was prepared via two methods. Prior to the final stage of sample synthesis, the compound's melting profile was determined using the Differential Scanning Calorimetry (DSC) measurement. In the first technique, the crushed AlPt powder was filled in an alumina tube and then sealed inside a tantalum tube. Finally, the whole assembly was homogeneously heated in argon atmosphere up to 1600 °C above the melting point of the compound and then cooled to 400 °C with a rate of 0.5 °C/min. Finally, the sample was annealed at 1000 °C for 6 days followed by a slow cooling of 1 °C/min to room temperature. In the second method, we followed the Bridgman-Stockbarger technique to prepare the AlPt sample. Here the crushed polycrystalline powder was filled in a custom designed sharp edged alumina tube, which was sealed in a tantalum tube. Then the whole assembly was heated up to 1600 °C and then slowly cooled to 1000 °C. The compositional analysis with energy-dispersive X-ray (EDX) spectroscopy shows stoichiometric Al and Pt, 1:1 composition in all the samples. Besides, the samples show similar pattern in the laboratory x-ray powder-diffraction (XRD). Here we used the AlPt samples prepared by both techniques for the ARPES measurements.

ARPES

Soft X-ray ARPES (SX-ARPES) measurements were performed at the SX-ARPES endstation [1] of the ADRESS beamline [2] at the Swiss Light Source, Switzerland, with a SPECS analyzer with an angular resolution of 0.07°. The photon energy varied from 350-1000 eV and the combined energy resolution was ranging between 50 meV to 150 meV. The temperature during sample cleaving and measurements was about 20 K and the pressure better than $1 \times 10^{-10}$ mbars. The increase of the photoelectron mean free path in the soft-X-ray energy range results, by the Heisenberg uncertainty principle, in a higher $k_z$ resolution of the ARPES experiment compared to measurements at lower photon energies [3], which was critical to measure the new Fermions in the bulk band structure of AlPt.

VUV-ARPES measurements were performed at the high-resolution- and nano-ARPES branch line of the beamline I05 at the Diamond Light Source, UK [4]. Measurements at the high-resolution branch were performed with a Scienta R4000 analyzer, and a photon energy range between 130 eV and 160 eV, at a temperature below 20 K. Measurements at the nano-ARPES branch were performed with a DA30 analyzer at a photon energy of 82 eV, a sample temperature below 25 K, and a beam spot size below 1 µm which allowed us to measure the Fermi surface map shown in Fig. 3C that originates from a single domain with a diameter of ~15 µm. Measurements in the VUV-ARPES regime are more surface sensitive than SX-ARPES and therefore most suitable to image the Fermi-arcs in AlPt.

Ab-initio calculations

We employed density functional theory (DFT) as implemented in the Vienna Ab Initio Simulation Package (VASP) [5,6] for the bulk calculations shown in Fig. 2 and the slab calculations shown Fig. 3B-D, as well as Wien2k [7] for the bulk calculations shown in Fig. 3A-B and Fig. S2.

For the VASP calculations, the exchange correlation term is described according to the Perdew-Burke-Ernzerhof (PBE) prescription together with projected augmented-wave pseudopotentials [8]. For the autoconsistent calculations, we used a 7x7x7 k-points mesh for the bulk and 4x4x1 for the slab calculations. The kinetic energy cut off was set to 400 eV. We calculated the surface states by using a slab geometry along the (001) direction. In order to achieve a negligible interaction between the surface states from both sides of the slab and reduce the overlap between top and bottom surface states, we considered a slab of 10-unit cells and 1 nm vacuum thickness. For the energy cuts, we used a 100x100 grid of K points. The spin resolved plots were done using PyProcar code [9,10].

The Wien2k calculations employed a full-potential linearized augmented plane-wave and local orbitals basis, as well as the PBE prescription of the exchange correlation term. The plane-wave cutoff parameter RMTKMAX was set to 7 and the irreducible Brillouin zone was sampled by 97,336 k-points. Spin-orbit coupling was included via a second variational procedure. The results from the Wien2k calculations were exported with the XCrySDen software [11] into the bxsf file format to plot the 3D Fermi surface with MATLAB.

The crystal structure and miller planes plotted in Fig. 1C were generated with VESTA [12].

Data availability

The data that support the plots within this paper and other findings of this study are available from the corresponding authors upon reasonable request.

Methods References

# Supplementary Materials

Outline

A. Photon energy dependent ARPES measurements

B. Complementary ARPES data taken at different cleavage planes and photon energies

C. Discussion of the CPGE and other transport probes in AlPt

D. Powder X-ray characterization of the AlPt samples

### A. Photon energy dependent ARPES measurements

To determine the momentum of the photoelectrons along the $k_z$ direction (the direction perpendicular to the cleavage plane), we employ the free-electron final state approximation with a fitting parameter $V_0$ called the inner potential, which allows us to convert the photoemission intensity from the measurement coordinate system with variable pair ($\theta$, $E_{kin}$) to momentum space ($k_y$, $k_z$) via the equations

$$k_y = \sqrt{2mE_{kin}/\hbar} \sin\theta, \quad (S1)$$

$$k_z = \sqrt{\frac{2m(E_{kin} + V)}{\hbar^2} - k_y^2}, \quad (S2)$$

where $\theta$ is the emission angle, $E_{kin}$ the kinetic energy of the electrons, m the electron mass, and $\hbar$ the reduced Planck constant. Since $V_0$ is a material dependent parameter, we performed photon energy dependent ARPES measurements over a wide range of photon energies to determine the appropriate inner potential by comparing the experimental Fermi surface in the $k_z$ vs. $k_y$ plane to the expected symmetry of the bulk Brillouin zone. Fig. S1A shows the simple cubic bulk Brillouin zone of AlPt (black solid lines) oriented with the (111) direction along $k_z$ (as is the case for our measurements presented in Fig. 2), as well as its boundaries (red solid lines) when slicing it along the M - Γ -M direction. Fig. S1B shows the experimental Fermi surface map obtained from a photon energy dependent measurement over a photon energy range hv= 355 – 1000 eV with an inner potential of $V_0$ = 26 eV. The orange and blue lines indicate the momentum positions of the measurements taken at hv=554 eV and 680 eV, which were chosen to probe the dispersion along the X – R – X and M - Γ -M directions, which are displayed in Fig. 2A(i) and 2E(i), respectively.

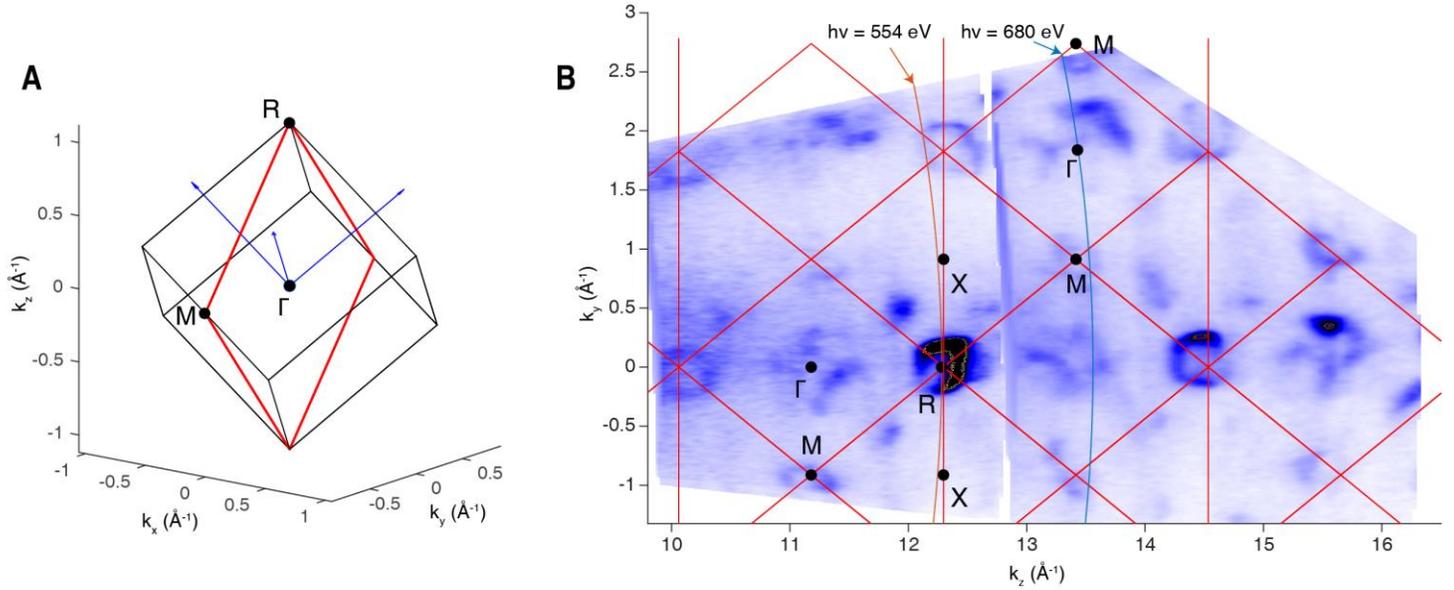

**Fig. S1. Photon energy dependent ARPES measurements.**
(**A**) 3D bulk Brillouin zone (solid black lines) oriented with the (111) direction along the $k_z$ axis. Brillouin zone boundary when slicing along the M - Γ -M direction is shown as red lines. The blue arrows indicate the reciprocal lattice vectors.
(**B**) Experimental Fermi surface map in the $k_z$ vs. $k_y$ plane taken over a photon energy range hv=355 eV – 1000 eV. Orange and blue lines indicate the momentum positions of the measurements taken at 554 eV and 680 eV.

B. <u>Complementary ARPES data taken at different cleavage planes and photon energies</u>

B1 <u>Additional SX-ARPES measurements taken on the (001) and (110) cleavage planes</u>

For completeness we show in Fig. S2 complementary measurements of the fourfold and sixfold fermions taken on the (001) and (110) cleavage planes. Fig. S2A shows the three-dimensional Fermi surface obtained by ab-initio calculations, as well as the measurement plane (shown as a red plane that is cutting through the Brillouin zone center) of a Fermi surface map taken on a strongly tilted (001) cleavage plane. Fig. S2B shows a Fermi surface map that is corresponding approximately to the same measurement plane as shown in Fig. S2A. Fig. S2C(i) shows a cut through the center of the Fermi surface map (indicated by the dashed magenta line in Fig. S2B) showing very similar dispersion as the one observed for the fourfold fermion shown in Fig. 2E(i), and which agrees well with the ab-initio calculation along a path cutting through the Γ point under a similar angle, which is shown in Fig. 2C(ii).

To better understand our data from the (110) surface, we show AlPt's Brillouin zone in Fig. S2D which is oriented with the (110) direction along the $k_z$ axis, and indicate the Brillouin zone boundaries by blue and red lines in the $k_x$ vs. $k_y$ and $k_z$ vs. $k_y$ planes, respectively. Our photon energy dependent measurement of the Fermi surface map in the $k_z$ vs. $k_y$ plane are shown in Fig. S2E, which agrees very well with the expected Brillouin zone periodicity, also indicating the momentum position of the measurements taken with hv=690 eV, which is very close to the plane containing M – R – M high symmetry points. Fig. S2F shows the Fermi surface taken at hv = 690 eV and Fig. S2G shows the corresponding line cut along $\bar{\bar{M}}$ - $\bar{R}$ -$\bar{\bar{M}}$ indicated by the magenta line in Fig. S2E, which also displays the sixfold fermions that were already shown in Fig. 2A(i). The slight loss of spectral intensity at the crossing point $\bar{R}$ is most likely due to a small misalignment of the sample or inaccurate choice of photon energy, such that the R point is slightly missed. However, the overall agreement with the ab-initio calculation for the M – R – M shown as solid blue lines in Fig. S2G, as well as the data already shown in Fig. 2, further corroborates the existence of the fourfold and sixfold crossing at the Γ and R point, respectively.

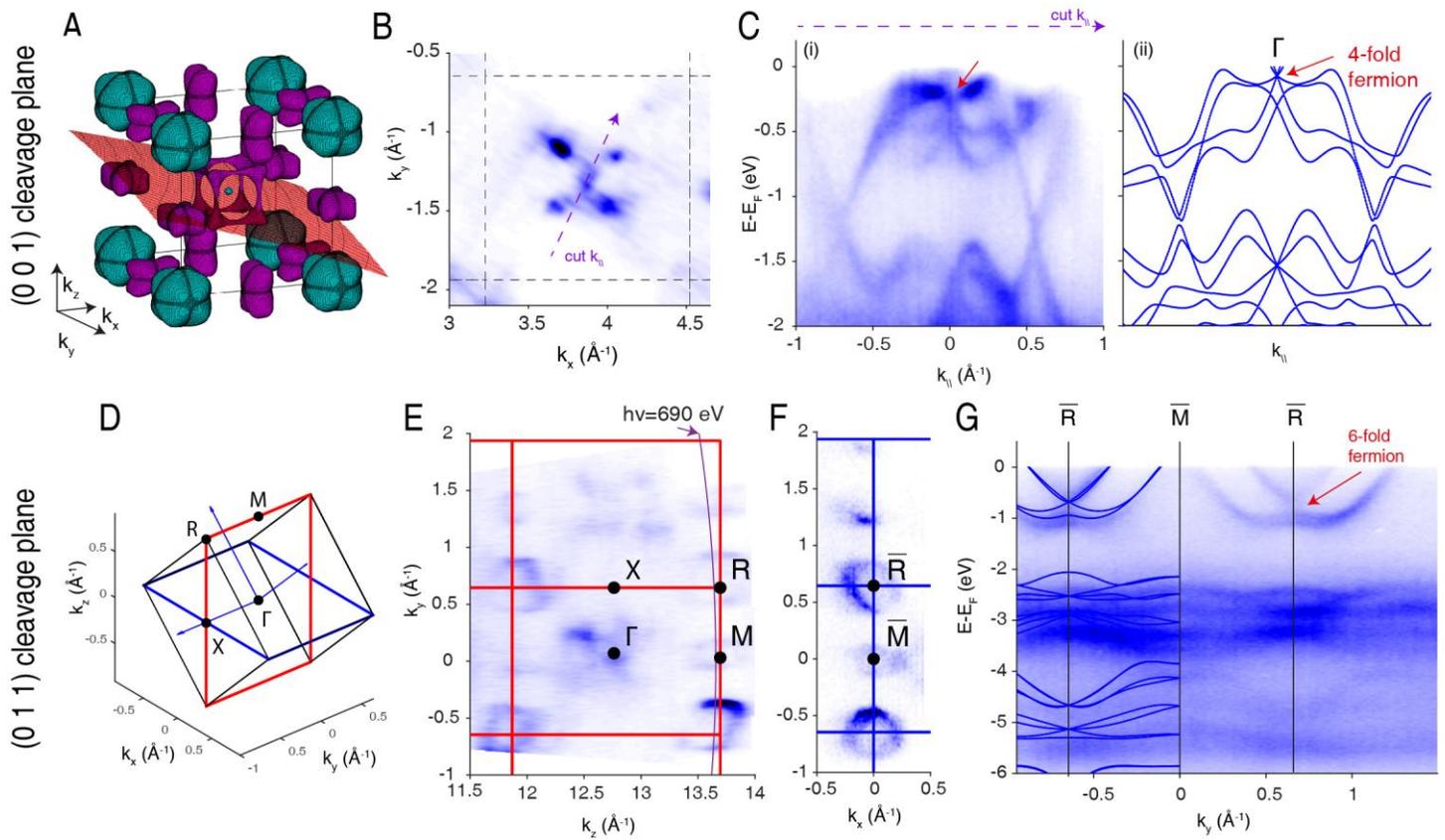

**Fig. S2 Complementary SX-ARPES data from the (001) and (110) cleavage planes.**
(**A**) 3D Fermi surface (electron pockets in green and hole pockets in magenta), as well as measurement plane for a tilted sample cleaved on the (001) surface (shown in red).
(**B**) Corresponding experimental Fermi surface map of a strongly tilted sample cleaved on the (001) surface, which was measured with hv=554 eV and right circular polarization.
(**C** (i)) Experimental dispersion along the direction indicated by the magenta dashed arrow in (**B**), measured with same photon energy and polarization.
(**C** (ii)) Calculated dispersion that is cutting through Γ under a similar angle as the dispersion shown in (**C** (i)) .
(**D**) Brillouin zone of AlPt oriented with the (110) direction along the $k_z$ axis. Its boundaries are indicated by blue and red lines in the $k_x$ vs. $k_y$ and $k_z$ vs. $k_y$ planes, respectively, and the blue arrows are the reciprocal lattice vectors.
(**E**) Photon energy dependent measurement of the Fermi surface map in the $k_z$ vs. $k_y$ plane, taken with right-circular polarization. Brillouin zone boundaries are shown as red dashed lines, and the magenta line shows the momentum position of the line cut taken at hv=690 eV.
(**F**) $k_x$ vs. $k_y$ Fermi surface map taken at hv=690 eV with the same polarization.
(**G**) Line cut measured at hv=690 eV corresponding to the magenta line shown in (**E**), which is very close to the R – M – R high symmetry direction. The corresponding ab-initio calculation along the R – M – R direction is shown as solid blue lines.

B2 Additional VUV-ARPES measurements taken on the (001) cleavage plane

To further illustrate the two-dimensional character of the surface Fermi-arcs measured on the (001) surface, we compare in Fig. S3 a Fermi surface taken on the (001) surface for three different photon energies hν = 130 eV, 145 eV, and 160 eV, which corresponds to probing the surface band structure at different momenta along the $k_z$ direction perpendicular to the sample surface. Although the spectral intensity of the Fermi surface arcs (FS arcs) is changing due to their photon energy dependent photoemission cross-section, the in-plane dispersion of the FS arcs is always the same in all three Fermi surface maps, which is the hallmark of 2D surface states that are nondispersive along $k_z$. We further replicated our results on a second sample that was measured with nano-ARPES at Diamond, and which also clearly shows the same S-shaped surface Fermi-arc dispersion (displayed in Fig. S4) at a different photon energy hν = 82 eV.

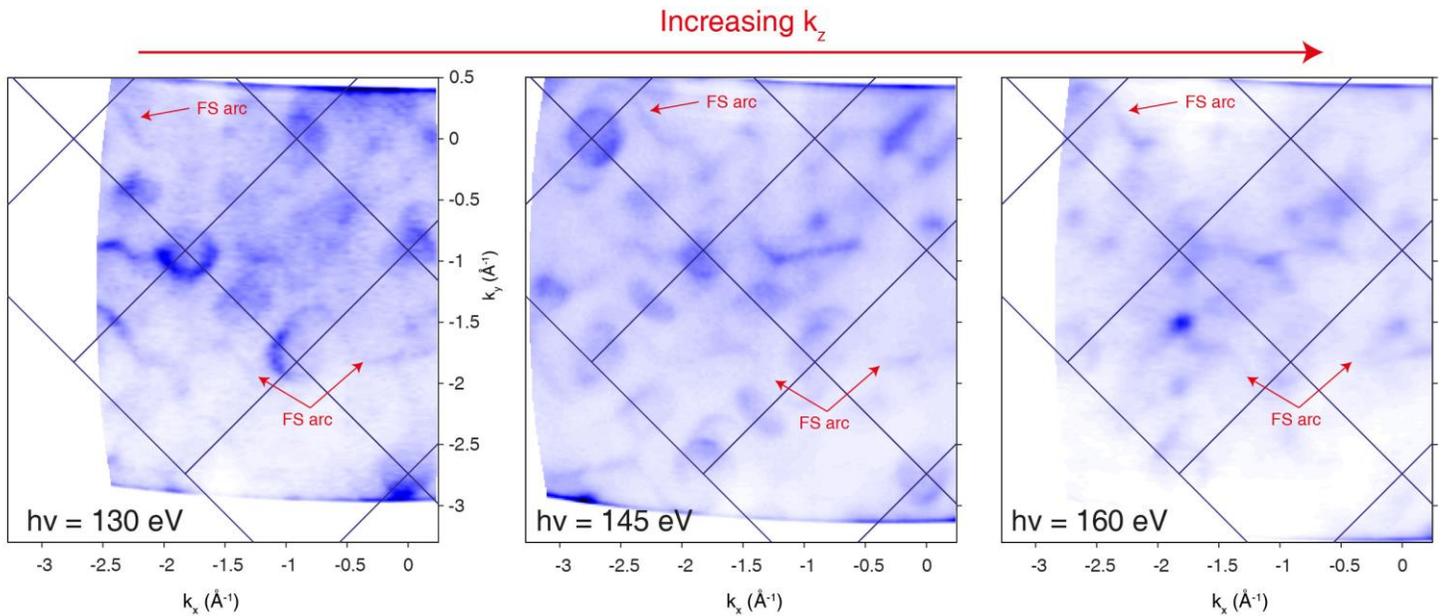

**Fig. S3. Fermi surface maps of the (001) surface for different photon energies**
Red arrows indicate the position of Fermi surface arcs which show the same in-plane dispersion for all three photon energies.

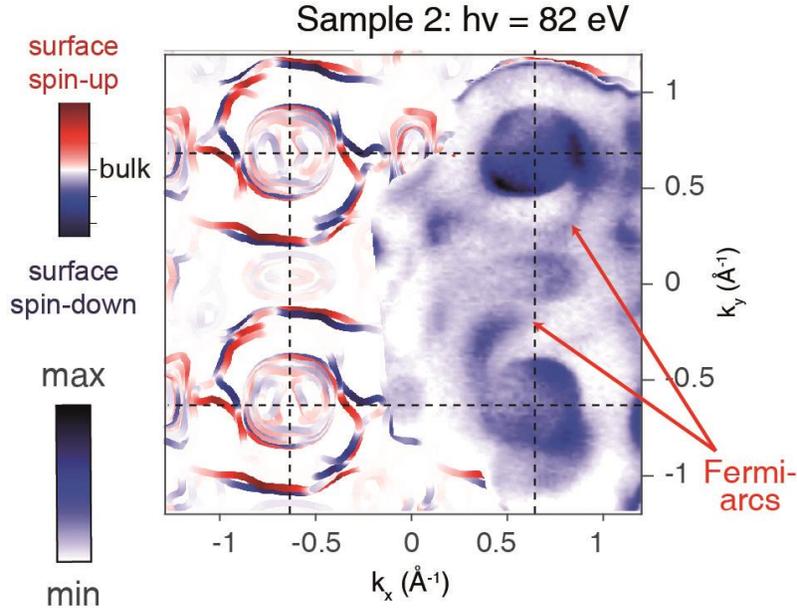

**Fig. S4.** Experimental Fermi surface map of the (001) surface for different sample at different photon energy hv=82 eV, replicating the appearance of surface Fermi-arcs. Ab-initio slab calculation as comparison on the left.

C. Discussion of the CPGE and other transport probes in AlPt

One may ask whether AlPt may be a good candidate for the investigation with other probes, such as transport or optics. To answer this question, it should be first noted that those two probes require different conditions for the response to be dominated by the topological carriers. DC transport probes ideally require that there only topological bands within a narrow range of energies around the Fermi level. Optical probes like the circular photogalvanic effect (CPGE), on the other hand, require that only topological bands are connected by resonant transitions at the applied frequency. The existence of bulk pockets at the Fermi level is not relevant for this effect as long as there are no transitions available to them at the applied frequency. Since we measure large bulk pockets for AlPt, this precludes any type of DC transport measurements that involve only the nodes. Removing this bulk pockets completely by alloying or doping might be possible but it requires significant changes in the band structure. Regarding optical probes, however, a frequency range dominated by the response at the Gamma point might be achieved in the current sample, for instance for the CPGE, as is illustrated in Fig. R3. Within the current DFT prediction, transitions between the lower and upper parts of the fourfold fermion at Gamma become active for the entire angular range for frequencies larger than ~130 meV, which contributes a quantized response in units of $e^3/h^2$ to the CPGE due to its Chern number $|C|=4$. At this frequency, there are two other regions in the Brillouin zone with allowed transitions that could provide unquantized contributions in the same energy range, one region is in the vicinity of R and another in the vicinity of M. However, we can see that that along other directions (such as along M-R) contributions from these nodes are not allowed in this energy range, and there is therefore only a limited angular range over which these transitions may contribute, which may be neglected in a first approximation.

The previous discussion illustrates that a measurement of an approximately quantized CPGE dominated by the fourfold node at G might be possible with the current samples for excitation frequencies above ~130 meV based on existing DFT calculations, which seem to agree well with our ARPES experiments. However, a more realistic calculation to determine the precise energy window over which a quantized response can be expected would also require knowledge about the unoccupied states, which are currently not accessible with ARPES and which would require further confirmation with other experimental probes e.g. with pump-probe ARPES. In addition, it is worth stressing that AlPt belongs to a large family of isostructural materials in SG 198, which means that many possibilities arise for alloying with different materials, for instance, alloying AlPt with Mg can drastically change its band structure and move the sixfold Fermion at R to the Fermi level. Taking AlPt as the only starting point that is currently known to conclusively possess Fermi arcs that are a fingerprint of their nonzero Chern number, it might be possible to obtain materials with related band structures where the optical response might have a larger frequency range dominated by the nodes, and the same might even be true for DC transport.

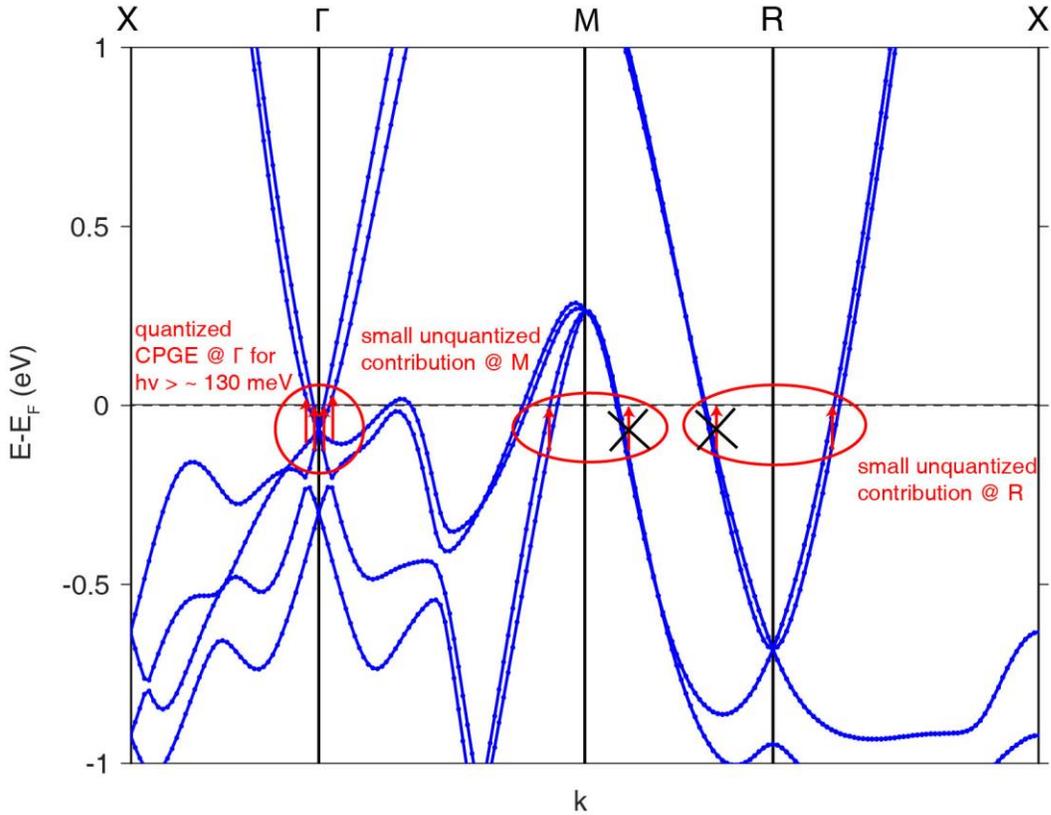

**Fig. S5: Energy windows for allowed optical interband transitions that contribute to the circular photogalvanic effect (CPGE).** Red arrows indicate allowed transitions, and crossed out red arrows indicate forbidden transitions. Contributions at G are allowed for the entire angular range for hv>130 meV and are quantized in units of universal constants $e^3/h^2$. Unquantized contributions from interband transitions in the vicinity of M and R are small due to the narrow angular range in which these transitions are allowed.

D. Powder X-ray diffraction measurements

Rietveld refinement of the powder X-ray diffraction (XRD) pattern of the AlPt sample, shown in Fig. S4, indicates that the sample crystallizes in the cubic chiral space group $P2_13$ (198). However, a small impurity phase of $Al_3Pt_5$ and $Al_{21}Pt_8$ has been detected in the XRD which we have highlighted as the additional phase. The two additional phases crystallize in orthorhombic P*bnm* (55) for $Al_3Pt_5$ and tetragonal $I4_1/a$ in $Al_{21}Pt_5$ (88). However, the symmetry of the Fermi surface maps and the excellent agreement with the ab-initio calculations clearly shows that the measured ARPES data reflects the majority phase of cubic AlPt.

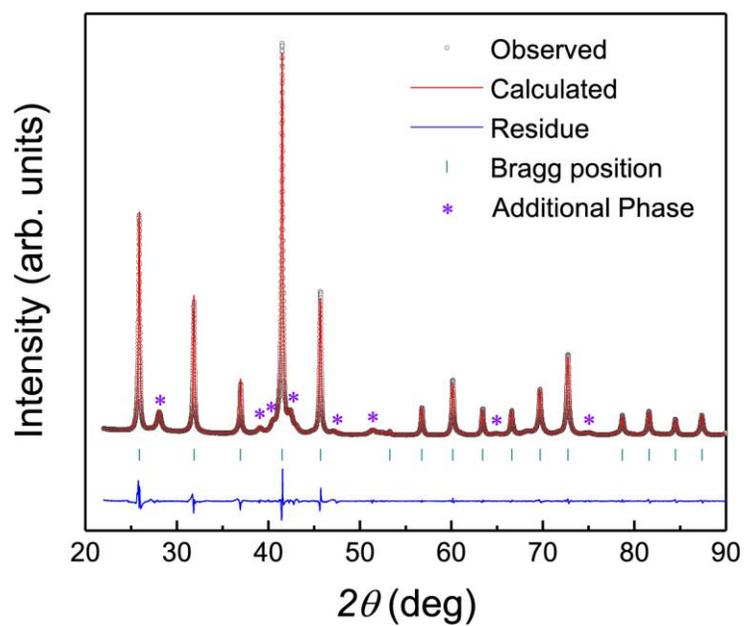

**Fig. S6. Powder X-ray diffraction pattern and Rietveld refinement**
The Retvield refinement indicates the presence of a majority phase of AlPt that crystallizes in the chiral space group 198, and a small minority phase of $Al_3Pt_5$ and $Al_{21}Pt_8$.